\documentclass[draft]{agujournal2019}

\graphicspath{{figures/}{pictures/}{images/}{./}}

\usepackage{siunitx}
\usepackage{hyperref}
\usepackage[colorinlistoftodos]{todonotes}
\usepackage{natbib}
\usepackage{datetime}
\usepackage{subcaption}
\newdate{date}{06}{09}{2012}
\date{\displaydate{date}}

\setuptodonotes{fancyline}

\begin{document}

%
%


\title{
Charting the Realms of Mesoscale Cloud Organisation using Unsupervised Learning
}

%
%




\authors{L. Denby\affil{1,2}}
\affiliation{1}{Danish Meteorological Institute, Copenhagen, DK}
\affiliation{2}{University of Leeds, UK}


\correspondingauthor{Leif Denby}{lcd@dmi.dk, l.c.denby@leeds.ac.uk}
\begin{center}
\today
\end{center}




\begin{keypoints}
\item The internal representation of an unsupervised neural network forms a map of all possible states of trade wind shallow cloud organisation
\item Mesoscale organisation does affect shortwave albedo even when correcting for changes in cloud-fraction
\item The temporal evolution of organisation is captured through the continuum representation of the cloud organisation map
\end{keypoints}

%
%


\begin{abstract}
Quantifying the driving mechanisms and effect on Earth's energy budget, of
mesoscale shallow cloud organisation, remains difficult.
Partly because quantifying the atmosphere's organisational state through
objective means remains challenging.
We present the first map of the full continuum of convective organisation
states by extracting the manifold within an unsupervised neural networks's
internal representation.
On the manifold distinct organisational regimes, defined in prior work, sit as
waymarkers in this continuum.
Composition of reanalysis and observations onto the manifold, shows wind-speed
and water vapour concentration as key environmental characteristics varying
with organisation.
We show, for the first time, that mesoscale shallow cloud organisation produces
$\pm 1.4\%$ variations in albedo in addition to variations from cloud-fraction
changes alone\footnote{In the final stages of preparing this manuscript we
became aware of independent, related work by \cite{alinaghi2023shallow}}. 
We further demonstrate how the manifold's continuum representation captures the
temporal evolution of organisation.
By enabling study of states and transitions in organisation (in simulations and
observations) the presented technique paves the way for better representation
of shallow clouds in simulations of Earth's future climate.
\end{abstract}

\section{Introduction}
\label{introduction}

From satellite imagery of Earth it is immediately clear that clouds often organise
into spatial patterns.
Names have been given to the most prominent patterns we
recognise (e.g. fronts, cyclones, cellular cloud-decks, etc) and we use this
classification to study the impacts of clouds on Earth's weather and climate
through their precipitation and interaction with radiation and atmospheric
circulation. 

One particular form of clouds, shallow tradewind cumuli, have a profound
importance in Earth's climate system due to their ubiquity and
net cooling effect, stemming from these clouds reflecting more incoming
short-wave radiation from the sun compared to the outgoing long-wave radiation
they permit \citep{Bony2004}.
Differing predictions of how these clouds will respond to a
warming climate accounts for most of the variation in climate sensitivity
between climate models (\cite{BonyDufresne2005}, \cite{Webb2006},
\cite{Medeiros2008}, \cite{Vial2013}), which highlights the urgent need to
better understand how these clouds form and interact with their environment
(one of the World Climate Research Programme's Grand Science Challenges,
\cite{Bony2015}).

One particular aspect of shallow convective clouds that is still poorly
understood, is their mesoscale organisation, both in quantifying what regimes
occur, the driving mechanisms behind them and extent to which mesoscale cloud
organisation impacts Earth's climate.
This interest in convective organisation stems from the observation that in
high-resolution (Large Eddy) simulations, clouds cluster (self-aggregate) under
certain conditions \citep{wing2018convective,muller2022spontaneous}, which
results in a change in cloud-fraction and net radiation for the same
domain-mean state.
Given the necessarily coarse resolution of climate models ($O(\si{10}{km}$)),
neither the processes driving organisation nor the organisational regimes can be
explicitly resolved and so the behaviour of these shallow clouds (and their
radiative impact) must be parameterised.

In the context of the EUREC\textsuperscript{4}A field campaign
\citep{Stevens2021eurec4a} work by \cite{Stevens2019} [S19]
developed a set of four classifications for shallow cloud organisation by
manual examination of visual satellite imagery.
These classes were motivated by the physical processes expected important in
different regimes, and have since formed the framing for many studies
investigating the conditions under which different forms of organisation occur,
both in observations \citep[][]{bony2020sugar,Schulz2021} [B21, S21] and in
simulations \citep{Dauhut2023}, with particular focus on the transition from
small to larger isolated detraining shallow clouds
\citep{Narenpitak2021,Saffin2023}.

An alternative approach was taken by \cite{Denby2020} [D20] who developed an
unsupervised machine learning approach to autonomously discover the possible
states of mesoscale organisation without imposing specific classes. This
approach of unsupervised learning was also taken by
\cite{Kurihana2022ClimInfo,Kurihana_2022Autoenc,Kurihana_2022_AICCA} using
clustering to produce individual classes of cloud patterns.
Differently again, \cite{Janssens2021} [J21] utilised the framework of traditional metrics used
for measuring clouds and their organisation (rather imposing specific classes),
concluding that rather than occurring in isolated regimes, cloud organisation
exists in a continuum (at least when viewed through these metrics).

By building on D20 we demonstrate in this work how the
continuum of convective organisation states is captured as an emergent
property of the internal \textit{embedding space} representation learnt by a
neural network through unsupervised learning.
Specifically, we will extract the low-dimensional manifold within the
high-dimensional embedding space on which all possible states of convective
organisation lie, and explore this manifold through the metrics and classes of
[J21] and [S19].
Through composition of reanalysis and observations onto this manifold we
characterise environmental conditions vary with convective organisation, we
quantify the effect of organisation on radiation correcting for changes in
cloud-fraction, and further
demonstrate how transitions between organisational states can be studied with
the manifold.


\section{Methods}

The primary tool used in this work is a convolutional neural network
which takes as input a 2D image-tile containing cloud imagery
(derived here from satellite observations) and produces a point in a
high-dimensional \textit{embedding} space (here 100-dimensional, as in [D20]),
a so-called \textit{embedding vector}.
During training the neural network has learnt to place tiles with similar cloud
structures nearby in the embedding space.
This is achieved by training the network on contrastive tile triplets of similar and
dissimilar cloud-patterns (produced by sampling from satellite imagery both
spatially closely-overlapping \textit{anchor-neighbor} and randomly distributed
\textit{distant} tiles).

The image-tiles utilised here were generated by performing locally planar
projections of observations from the geostationary GOES-16 satellite to produce
$256 \times 256$ pixel input tiles spanning $\si{200}{km} \times \si{200}{km}$.
In extension to D20's use of truecolor RGB-composite tiles, we here also
generate a set of tiles using the three \textit{water-vapour window} infrared
channels (11, 14 \& 15) and train a separate IR-tiles model.
This enables characterisation of convective organisation throughout the diurnal
cycle (including nighttime), enabling analysis of cloud evolution over multiple
days (see \autoref{sec:trajectory-transitions}).

We use the Level-2 $\Delta x \approx \si{4}{km}$ 1-hourly CERES radiation products
derived from the GOES-16 geostationary satellite \citep{ASDC2018SatCORPS}
\footnote{\cite{MODIS-Syn1Deg1Hour-Terra-Aqua} produces similar but smaller
albedo variations due to its coarser resolution}.
Environmental characteristics associated with different states of organisation
were extracted by resampling ERA5 reanalysis \citep{ERA5} onto the dataset tiles.
For cloud-fraction and cloud metric calculations \citep[]{DenbyJanssens2022} we
use the shallow cloud-mask as defined by \cite{bony2020sugar}, thresholding on
GOES-16 channel 13 ($\si{10.35}{\mu m}$) brightness temperature ($\si{280}{K} <
T^{ch13}_b < \si{290}{K}$).
To aid comparison with prior work we have chosen to restrict our analysis to
tiles with minimum cloud top-temperature above freezing level $T_c >
\si{273}{K}$.

\section{Results}

\subsection{The manifold of mesoscale cloud organisation}
\label{sec:manifold-extraction}

\begin{figure}
    \begin{subfigure}[t]{0\textwidth}
        \phantomcaption
        \label{fig:manifold-with-insets}
    \end{subfigure}
    \begin{subfigure}[t]{0\textwidth}
        \phantomcaption
        \label{fig:manifold-cloud-metrics-cloud-size}
    \end{subfigure}
    \begin{subfigure}[t]{0\textwidth}
         \phantomcaption
        \label{fig:manifold-cloud-metrics-open-sky}
    \end{subfigure}
    \begin{subfigure}[t]{0\textwidth}
         \phantomcaption
        \label{fig:manifold-cloud-metrics-woi3-lwp}
    \end{subfigure}
    \begin{subfigure}[t]{0\textwidth}
         \phantomcaption
        \label{fig:manifold-cloud-metrics-std-cth}
    \end{subfigure}
    \begin{subfigure}[t]{0\textwidth}
         \phantomcaption
        \label{fig:manifold-sgff-class-distribution}
    \end{subfigure}
    \includegraphics[width=1.0\textwidth]{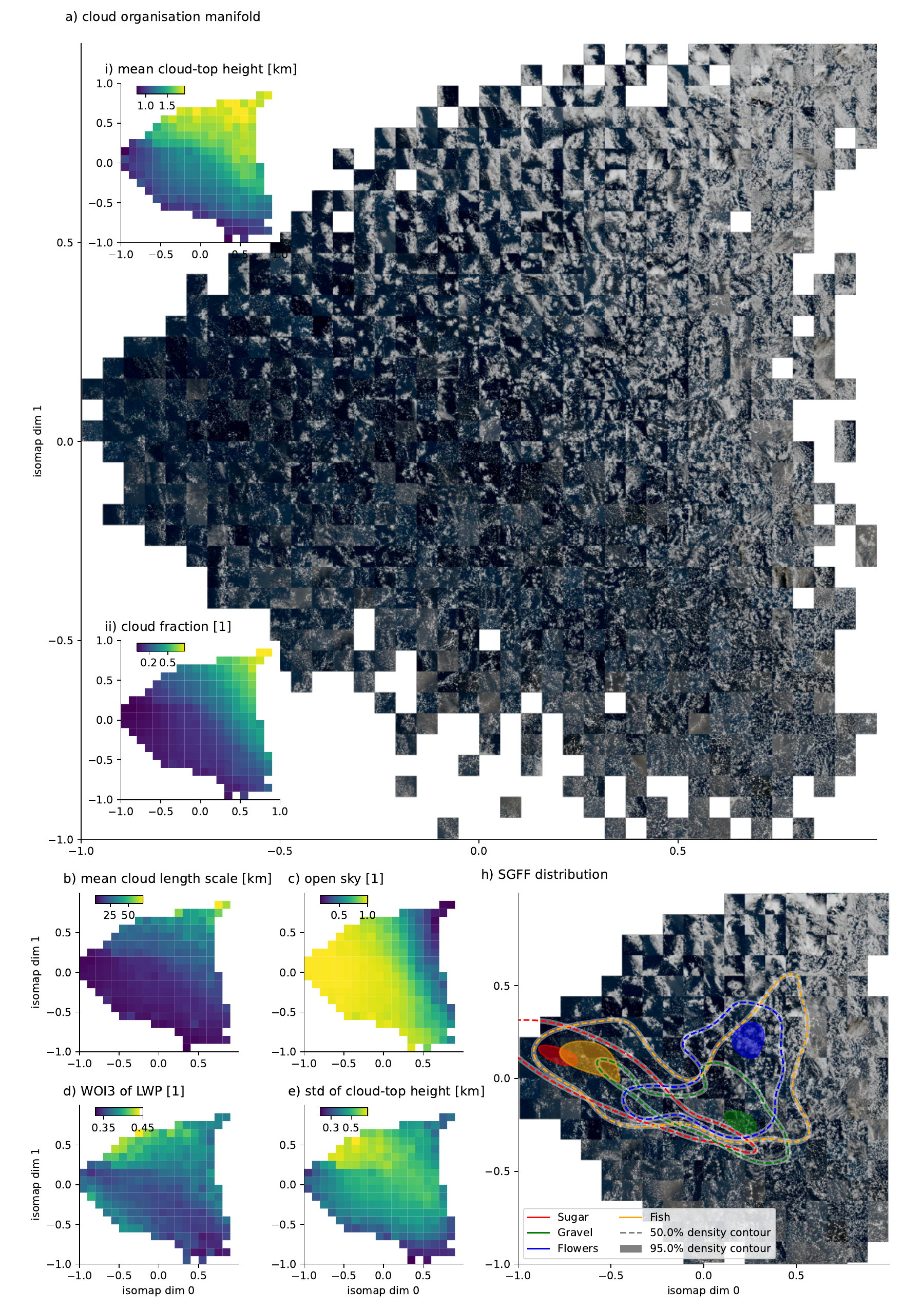}
    \caption{
        Convective organisation embedding manifold a) visualised by the
        anchor tile from the closest anchor-neighbour pair within a fixed interval
        across the manifold inset with i) mean cloud-top height and ii)
        cloud-fraction, together with conventional cloud metrics of J21 
        b) characteristic cloud-size ($L_c$), c) contiguous open-sky fraction ($f_{os}$), d)
        directional alignment of liquid water path ($a_{LWP}$), e) standard
        deviation of cloud-top height $\sigma_{CTH}$ and h) per-class
        distribution of S19 organisation classes.
    }
    \label{fig:embedding-manifold}
\end{figure}

In this section we will demonstrate how the continuum of cloud organisation
states can be extracted as an emergent property of the embedding space utilised
by a trained neural network.
We do this by examining the topological structure of the embeddings produced
by the neural network applied to a dataset of 10,000 triplet tiles.

As the neural network has learnt during training to place similar cloud
structures close in the embedding space, the occurrence of evolution between
organisation states will manifest as a continuum of points in the embedding
space (assuming a dataset large enough to contain the in-between states).
The extent to which this point-cloud maps out either isolated regions, clusters
connected by isolated paths, or full manifolds of smooth evolution, tells us not
only what kinds of regimes exist, but also how distinct they are and what
transitions between regimes are actually observed in nature.
Prior work by [S19] suggests that organisation in the tropical Atlantic comes
in four distinct forms, which would manifest as the embeddings lying in four
isolated clusters.
However, clustering analysis shows no evidence of isolated clusters in the
embedding space.
Indeed, should we expect the atmosphere to gravitate to only few
distinct forms of cloud organisation?
If the atmosphere is constantly evolving between different forms of
organisation, why would it stay "stuck" in specific isolated regimes of
organisation rather than spending as much time transitioning between regimes?
Since attempts to find isolated clusters failed, we moved to techniques which
are able to maintain the continuum representation that embedding vectors of
organisation provide.
By applying manifold extraction techniques to the embedding space we found that
the tile embeddings do indeed appear to lie on a low-dimensional manifold
within the high-dimensional embedding space.

We applied the Isomap manifold extraction method \citep{Tenenbaum_2000} to
transform the 100D embedding-vector point cloud into a 2D plane, the result of
which is visualised in \autoref{fig:manifold-with-insets} by rendering tiles
for individual points across the manifold.
Isomap was selected because it through construction of a nearest-neighbour
graph through the entire tile embedding point-cloud, ensures that the topology
of the manifold is unchanged (i.e. does not introduce or remove new paths
between points).
The manifold extraction allows us to perform dimensionality reduction
of the embedding space, by extracting only the part of the embedding
space that the neural network has actually utilised.
Using a manifold extraction method, rather than for example Principal Component
Analysis (PCA), avoids assuming that the embedding points lie on a
high-dimensional plane since Isomap follows the curvature of the manifold
spanned by the tile embedding points.

To visualise the embedding manifold we select the closest
anchor-neighbor pair for each fixed-width bin across the embedding manifold,
and render the anchor tile from that pair.
This produces tiles with a clearly discernable pattern by exploiting that where
the trained neural network is unable to place two anchor-neighbor tiles in
close proximity (in the embedding space), this suggests that either a) these
tiles are very different in the cloud structures present (unlikely given the
spatial overlap of anchor-neighbor tiles) or b) it is not possible for the
neural network to characterise the organisational state of these two tiles and
so these tiles are not in a clearly descernable organisational state.
Conversely, closely-spaced anchor-neighbor tiles are not only similar, but
in an organisational state clearly identifiable by the trained neural network.

We can by eye immediately identify differences in the morphological features of
clouds in different parts of the manifold,
with smaller isolated clouds concentrated in the lower left, larger isolated
clouds in the upper corner and the bottom right populated by cellular cloud
structures (cloud-size and cloud-fraction \autoref{fig:manifold-with-insets}
insets i and ii respectively).
Having produced this 2D "map" of cloud organisation through extraction of the
underlying manifold utilised by the neural network, 
we next to turn to examining the organisational states spanned within this
manifold, both in terms of contemporary means of cloud metrics and organisation
classes identified by J21 and S19 respectively, and later through quantifying
the environmental conditions associated with different forms of organisation
with and radiation effects of organisation.

This is done by aggregating observations and reanalysis datasets (coincident in
time and space with the sampled tiles) onto the 2D manifold by binning each
physical variable in turn by the 2D manifold dimensions.

\subsubsection{Cloud-metrics on manifold}

We next set the extracted manifold in a more familiar context by computing the
set of four cloud property metrics identified by J21 as collectively being able
to describe the majority of variation between organisational regimes: b) the
cloud length-scale ($L_c$), c) open-sky amount
\citep[$f_{os}$,][]{Antonissen2019}, d) directional anisotropy ($WOI_3$) of
liquid water path \citep[$a_{LWP}$,][]{Brune2021} and e) standard deviation of
cloud-top height ($\sigma_{CTH}$).
The variation of these metrics across the embedding manifold is shown in the
respective subfigures of \autoref{fig:embedding-manifold}, which depict
the mean value for each variable across all tiles falling within a given bin in
on 2D manifold.

Visually, it is immediately clear that using just these metrics there are many
ways to split up the embedding manifold into regions of similar
characteristics, with both smooth and abrupt changes in the cloud property
metrics across the manifold.
As cloud-size is a very familiar characteristic that relates to how a
cloud was formed and the radiative impact it has, we will concentrate on
separately examining the large regions of the manifold where the characteristic
cloud-size remains near-constant.
For the very smallest clouds ($L_c < \si{20}{km}$ in
\autoref{fig:manifold-cloud-metrics-cloud-size}) the principal variation is
the decrease in the characteristic open-sky amount
(\autoref{fig:manifold-cloud-metrics-open-sky}) with increasing cloud-top
height (\autoref{fig:manifold-with-insets} inset i) (and lesser increase in
cloud-fraction, \autoref{fig:manifold-with-insets} inset ii), which
describes the transition from scattered isolated clouds (left in the manifold)
to more cellular (cold-pool arc) cloud formations (bottom right of the
manifold).
The largest clouds ($L_c > \si{20}{km}$) are found in more varied configurations;
specifically, with either a) very variable cloud-top height
(\autoref{fig:manifold-cloud-metrics-std-cth}), intermediate cloud-fraction and
large regions of open-sky (the large isolated clouds in the top left of the
manifold), but also situations b) with less variation in cloud-top height and
lower cloud-fraction (the top right of the manifold).

It is worth pausing here to consider how a 2D plane appears to be adequate to
map all regimes of the shallow cumulus organisation when J21 showed that a full
set off four metrics are needed.
We, conjecture that this is principally because J21 used PCA and so explicitly
assumed that variations in mesoscale organisation can be described by a set of
linear basis defined by the metrics used.
However, as noted by J21 it is not clear that the linear
decomposition is physically meaningful for cloud organisation, evidenced by the
metrics not being linearly separable.
Said differently, it is quite possible that the full set of four metrics are only
required to distinguish a subset of organisational states and for other regimes
the metrics actually covary.
We see this behaviour in our analysis as the first two metrics ($L_c$ and $f_{os}$)
show an unidirectional gradient across the entire embedding
manifold, whereas the last two metrics, $a_{LWP}$ and $\sigma_{CTH}$
(\autoref{fig:manifold-cloud-metrics-woi3-lwp} and
\ref{fig:manifold-cloud-metrics-std-cth}) primarily
vary within smaller regions where the first two metrics are near-constant.

The extent to which the space of possible organisational states is indeed
inherently two-dimensional, will be investigated using topological analysis
tools in future work.
However, through the compositing cloud and environment characteristics (the
latter in \autoref{sec:manifold-environment}) on the 2D embedding manifold we do find physically
meaningful interpretations of variability observed and so we expect that to
leading order the 2D manifold here is a useful tool to understanding what kinds
of organisation form and why they form.


\subsubsection{Finding Sugar, Fish, Flowers and Gravel}
\label{sec:finding-sgff}

We next turn to examining the extent to which the convective organisation
classes (\textit{Sugar, Gravel, Fish} and \textit{Flowers}) described by B19
appear as distinct regions on the embedding manifold, examining
whether the self-supervised neural network has "discovered" these manually
defined classes of organisation.
This is done by producing embeddings with the trained neural network at the
same locations which have been manually labelled as belonging
to one of the four classes and, for each of the four classes in turn, plotting
the distribution of these embeddings over the manifold (see
\autoref{fig:manifold-sgff-class-distribution}), thereby showing where
the trained neural network would place a tile with a given manual
classification.
The manually-labelled dataset produced for the
EUREC4A field campaign \cite{Stevens2021eurec4a} by \cite{Schulz2022} which gives at
$0.01^{\circ}\times 0.01^{\circ}$ resolution for 1/7/2020 to 2/3/2022 in the
tropical Atlantic the number of people labelling a given location in the four
classes.
As \cite{Schulz2022} any location with over $60\%$ agreement in labelling is
designated as belonging to a specific class.
The per-class distribution is visualised by the $50\%$ density contour (showing
the spread) and then $90\%$ density contour (containing the peak) computed
using Kernel Density Estimation (KDE).


The distribution of tiles labelled as \textit{Sugar}, \textit{Gravel} and
\textit{Flowers} peak in distinct regions of the embedding manifold suggesting
these three classes of organisation have distinctive characteristics
identifiable using unsupervised learning (in contrast to \textit{Fish},
discussed below).
The spatial regions labelled \textit{Sugar} are concentrated in the region of the
embedding space where scattered isolated cumuli with low vertical extent are
concentrated, \textit{Gravel} is associated with (often multi-) cellular cloud
structures formed by cold-pool arcs, and \textit{Flowers} in the region of
larger, deeper isolated cumuli,
all in agreement with B19.

However, although the distributions of these three classes peak in isolated
regions in embedding manifold, they also show a broad and varying overlap
between organisation classes.
The \textit{Sugar} class appears the most separated from the rest, fitting with
it being the archetypical organisation comprising randomly scattered small
shallow cumuli.
\textit{Gravel} organisation extends into the region of \textit{Sugar}, fitting
with \textit{Gravel} being characterised by cloud-free voids (driven by
cold-pools) within regions if shallow scattered cumuli.
For tiles labelled as \textit{Flower} organisation, the distribution extends to
encompass the peak of the \textit{Gravel} cloud distribution, which is
consistent with these larger isolated clouds in some cases being associated
with cold-pool arcs \citep{Cui2023}.
Finally, we consider the distribution of tile samples from regions labelled as
\textit{Fish} which is concentrated in same part of the embedding manifold as the
\textit{Sugar} class and extends to include all three of the other classes.
We conjecture that this is due to the large characteristic length-scale of
\textit{Fish} organisation (visually often O($\si{1000}{km}$)).
Specifically, on the length-scale of tile-size used in this work it appears
that \textit{Fish} organisation is comprised from smaller patterns of mesoscale
organisation rather than being distinct.

As mentioned above, the overlaps in the distribution across the embedding
manifold of regions manually labelled as belonging to separate kinds of
organisation can to some degree be explained by the fact that some kinds of
mesoscale cloud structures could be expected to occur together based on
physical mechanisms expected to be associated with their formation.
The isolated nature of the peaks of the manifold distribution of the three
\textit{Sugar, Gravel and Flower} classes above supports findings of
\citep{Stevens2019} that these kinds of organisation have distinct
characteristics.
However, looking across all tiles these same classes do not show up as
isolated distributions across the embedding manifold. Said another way, these three
classes of organisation should be considered as useful waymarkers in the full
continuous state-space of mesoscale convective organisation, rather than
defining the only distinct kinds of organisation to which most kinds of shallow
cumulus cloud formations belong.

\subsection{Environmental characteristics of mesoscale organisation}
\label{sec:manifold-environment}

\begin{figure}
    \begin{subfigure}[t]{0\textwidth}
        \phantomcaption
        \label{fig:manifold-environmental-conditions-cl-windspeed}
    \end{subfigure}
    \begin{subfigure}[t]{0\textwidth}
        \phantomcaption
        \label{fig:manifold-environmental-conditions-cl-moisture}
    \end{subfigure}
    \begin{subfigure}[t]{0\textwidth}
        \phantomcaption
        \label{fig:manifold-environmental-conditions-lts}
    \end{subfigure}
    \begin{subfigure}[t]{0\textwidth}
        \phantomcaption
        \label{fig:manifold-environmental-conditions-bl-windspeed}
    \end{subfigure}
    \begin{subfigure}[t]{0\textwidth}
        \phantomcaption
        \label{fig:manifold-environmental-conditions-bl-moisture}
    \end{subfigure}
    \begin{subfigure}[t]{0\textwidth}
        \phantomcaption
        \label{fig:manifold-environmental-conditions-sst}
    \end{subfigure}
    \includegraphics[width=\textwidth]{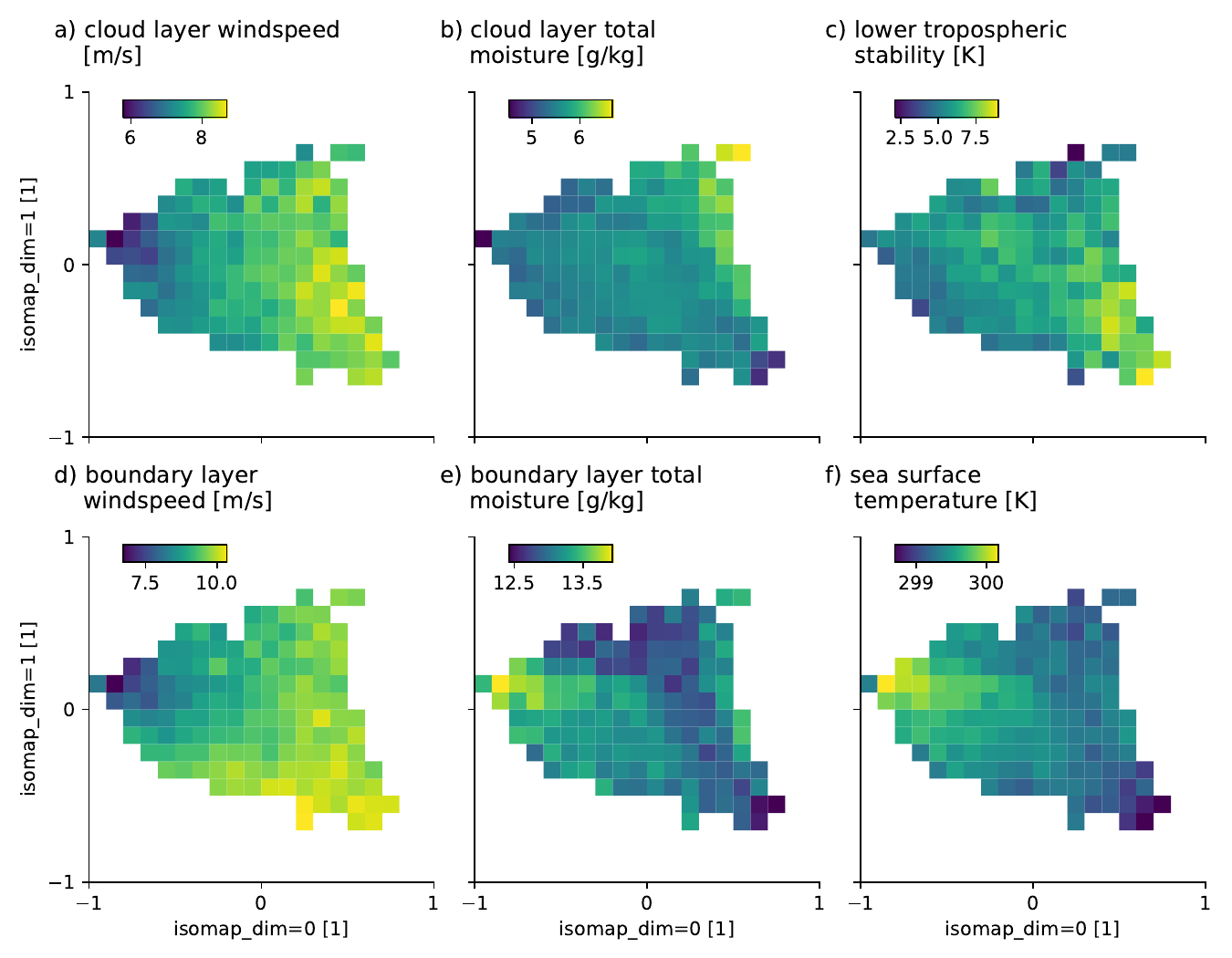}
    \caption{
        Bin-mean values of environmental characteristics derived from ERA5
        reanalysis across the embedding manifold
    }
    \label{fig:manifold-environmental-conditions}
\end{figure}

Next we will examine how the environment varies with different forms of
organisation by compositing ERA5 reanalysis on the embedding manifold.
Across the manifold we will quantify how organisation varies with environmental
characteristics (\autoref{fig:manifold-environmental-conditions}) and contrast
these relationships with findings in prior work.

In agreement with B21 and S21, in conditions with the lowest wind-speeds
(\autoref{fig:manifold-environmental-conditions-cl-windspeed} and
\ref{fig:manifold-environmental-conditions-cl-windspeed})
and relatively warm sea-surface temperatures
(\autoref{fig:manifold-environmental-conditions-sst}), we find the archetypical
scattered shallow cumuli and in conditions with lowest sea-surface temperature
and highest wind-speeds, the shallow more cellular cloud structures formed by
evaporation-driven cold-pools (these forms of organisation are also called
\textit{Sugar} and \textit{Gravel}, see \autoref{sec:finding-sgff}).
Contrasting the manifold regions with the smallest (and shallowest, $z_{CTH} <
\si{2}{km}$) and largest (and deepest, $z_{CTH} \approx \si{4}{km}$) isolated
clouds we find that the environmental factor which most clearly
differentiates these regions to be sub-cloud and cloud-level moisture
(\autoref{fig:manifold-environmental-conditions-bl-moisture} and
\ref{fig:manifold-environmental-conditions-cl-moisture}), with
higher cloud-level and lower sub-cloud moisture in the latter case.
In fact, the moisture and wind-speed appear nearly orthogonal in their
variation across the embedding manifold, so that for contours of
fixed moisture, the wind-speed uniquely defines a location on the manifold
and thus the kind of organisation typically associated with these conditions.
In contrast to B21 and S21, we do not find the strongest stability conditions
(\autoref{fig:manifold-environmental-conditions-lts}) to be associated with
large and deep isolated clouds, but rather with the strongest degree of
cellular organisation.

\subsection{Radiative impact of mesoscale organisation}

\begin{figure}
    \begin{subfigure}[t]{0\textwidth}
        \phantomcaption
        \label{fig:isomap-radiation-sw-albedo}
    \end{subfigure}
    \begin{subfigure}[t]{0\textwidth}
        \phantomcaption
        \label{fig:isomap-radiation-cf-sw-albedo-fit}
    \end{subfigure}
    \begin{subfigure}[t]{0\textwidth}
        \phantomcaption
        \label{fig:isomap-radiation-cf-sw-albedo-model-missfit}
    \end{subfigure}
    \begin{subfigure}[t]{0\textwidth}
        \phantomcaption
        \label{fig:isomap-radiation-cloud-optical-depth}
    \end{subfigure}
    \begin{subfigure}[t]{0\textwidth}
        \phantomcaption
        \label{fig:isomap-radiation-cloud-particle-radius}
    \end{subfigure}
    \begin{subfigure}[t]{0\textwidth}
        \phantomcaption
        \label{fig:isomap-radiation-lwp}
    \end{subfigure}
    \includegraphics[width=\textwidth]{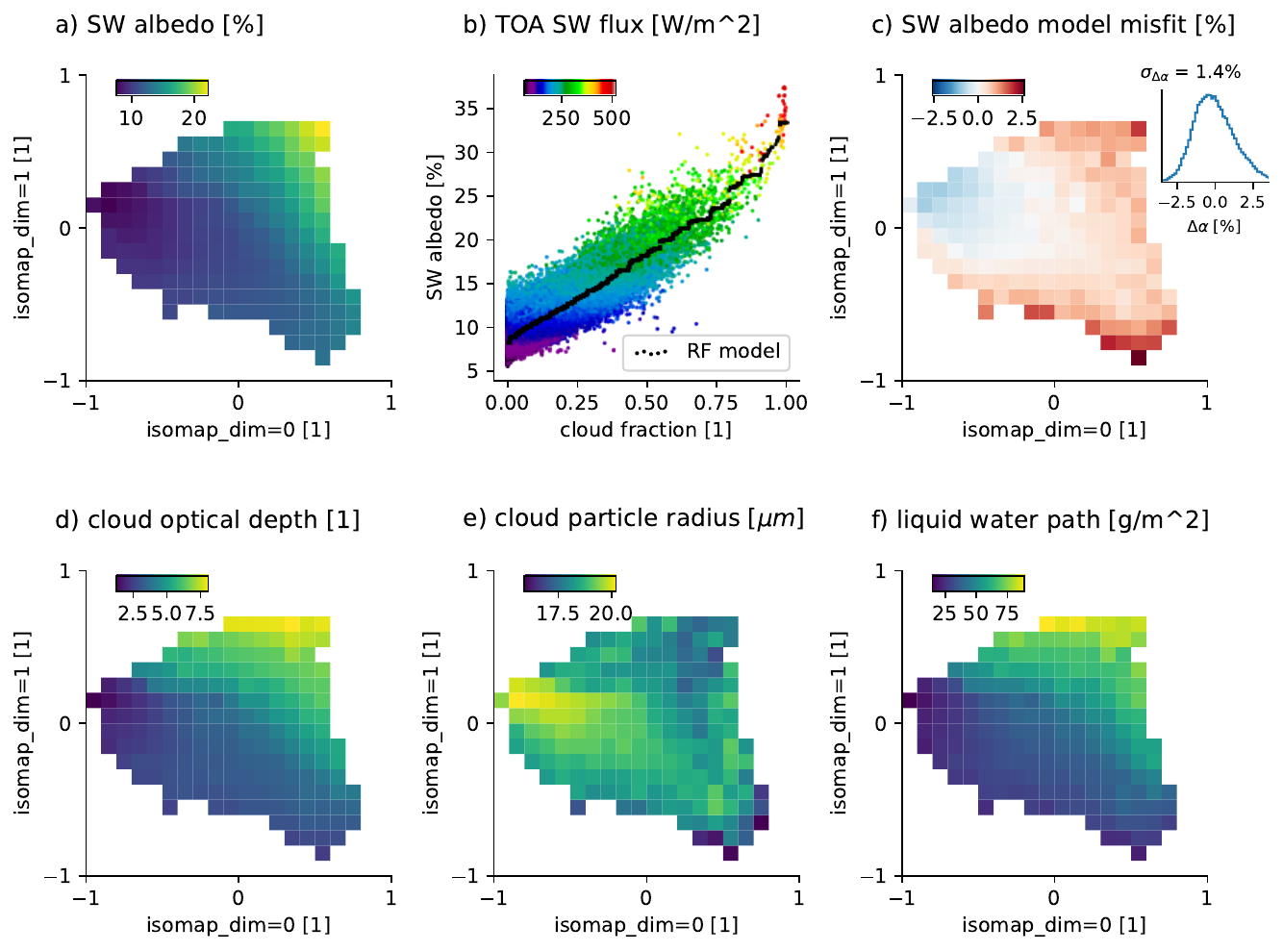}
    \caption{
        Variation of shortwave albedo a) across the embedding manifold, with
        cloud-fraction based model miss-fit both b) point-wise and c) across
        the manifold (distribution in inset) showing the effect of organisation
        alone (discarding cloud-fraction) together with cloud properties
        affecting albedo d), e) and f)
    }
    \label{fig:isomap-radiation}
\end{figure}

The principal reason for being interested in convective organisation is the
possible impact of organisation of shallow convective clouds on Earth's
radiation budget.
Using the four classes of organisation by S19, B21 concluded that \textit{"for a
given low-cloud amount [we] did not find a significant effect of cloud
organisation on the shortwave albedo."}
From B21 it is clear that to leading-order that mesocale albedo is controlled by
cloud-fraction, however, as the four regimes of organisation studied show
little overlap in cloud-fraction, we wonder strong the effect of organisation
is for a given cloud-fraction and so attempt to separate the effect of
cloud-fraction and organisation on albedo.

We do this by first fitting a simple model (a random forest, in effect an
optimal-binning algorithm producing step-wise predictions) to predict the
tile-mean shortwave albedo (\autoref{fig:isomap-radiation-sw-albedo}) from the tile
cloud-fraction.
The result of this fit can be seen in
\autoref{fig:isomap-radiation-cf-sw-albedo-fit}) and shows the same monotonic
increase in albedo with cloud-fraction as B21.
The figure also highlights the significant spread in albedo for a given
cloud-fraction.
We next plot the mean error of this simple model across the embedding manifold
(\autoref{fig:isomap-radiation-cf-sw-albedo-model-missfit}), thereby getting a
direct visualisation of the extent to which organisation alone affects albedo.
In contrast to B21, we do find that organisation effects the shortwave albedo
besides through the change of cloud-fraction, with the model albedo miss-fit of
$\sigma_{\Delta \alpha} \approx 1.4\%$.

To set the effect of organisation on radiation in context we have included in
\autoref{fig:isomap-radiation-cloud-optical-depth},
\ref{fig:isomap-radiation-cloud-particle-radius} and
\ref{fig:isomap-radiation-lwp}, the mean optical depth, cloud particle radius
and liquid water path respectively.
Organisation dominated by small scattered cumuli has an anomalously low albedo
(compared to what would be predicted from cloud-fraction alone) appearing to
result from low cloud optical depth, which in turn results from larger drops and
lower total amount of liquid water than other states of organisation across the
manifold.
In contrast, regimes with higher albedo generally have smaller droplets
and higher liquid water path (in cellular organisation) or very small droplets
and high liquid water path (large isolated cumuli).

The assumption that cloud organisation only effects the amount of reflected
shortwave through changes in cloud-fraction, is in effect assuming that all
shallow convective cloudy pixels have the same albedo and the only important
factor is how many pixels contain cloud.
We have shown here that properties of the cloudy pixels do indeed effect
albedo, and the connection between mesoscale organisation and reflected
shortwave radiation is more than just cloud-fraction. 
To fully unpick the role of cloud-properties changing with organisation, further
work should be done with in-situ observations and modelling studies to
compliment the remote-sensing retrievals used here.

\subsection{Mapping the temporal evolution of organisation}
\label{sec:trajectory-transitions}

\begin{figure}
    \begin{subfigure}[t]{0\textwidth}
        \phantomcaption
        \label{fig:isomap-2nd-feb-trajectory-manifold}
    \end{subfigure}
    \begin{subfigure}[t]{0\textwidth}
        \phantomcaption
        \label{fig:isomap-2nd-feb-trajectory-samples}
    \end{subfigure}
    \includegraphics[width=\textwidth]{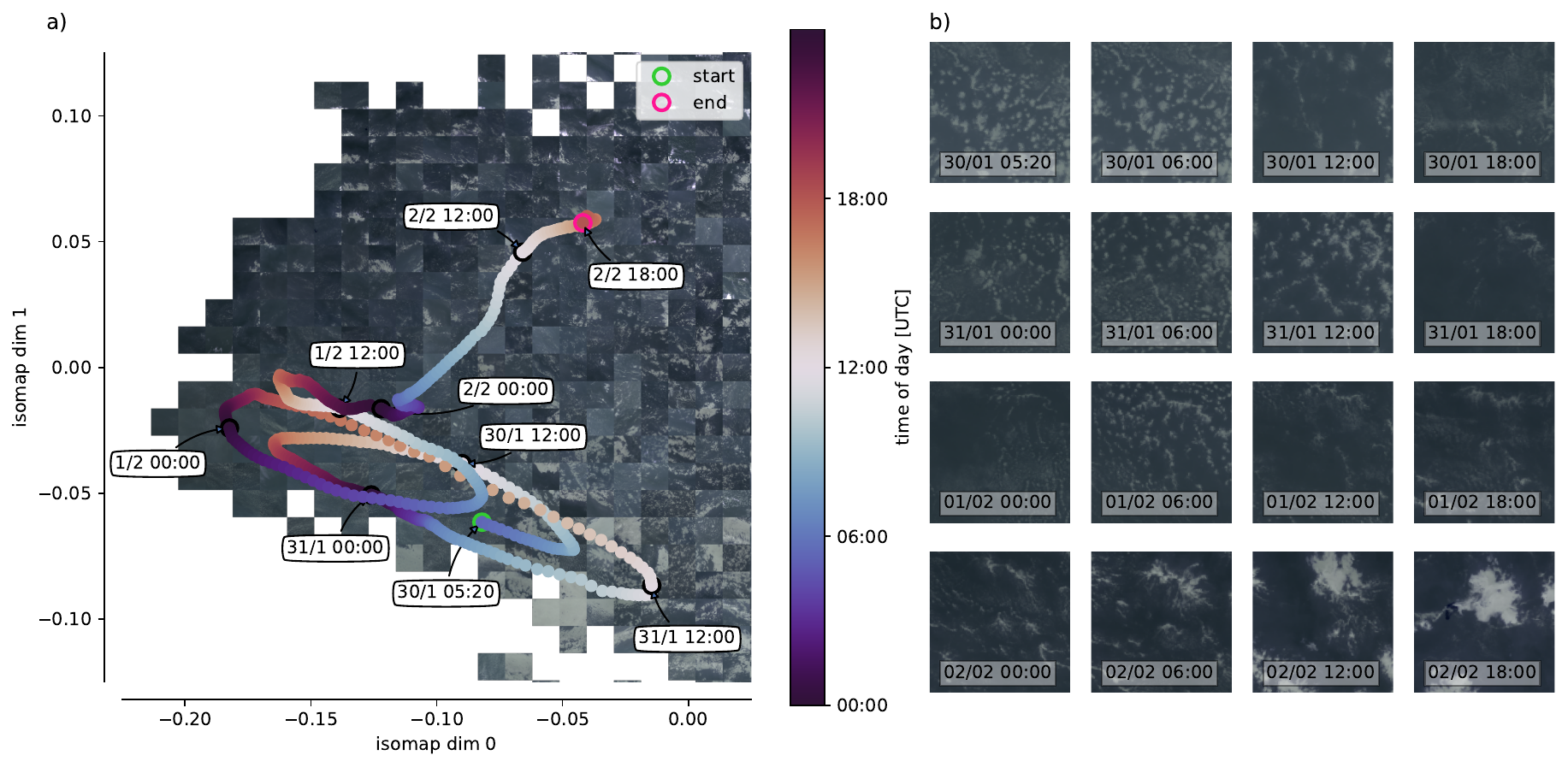}
    \caption{
        Evolution of cloud organisation along 4-day Lagrangian trajectory
        arriving at Barbados on 2nd Feb 2020 visualised a) on the embedding
        manifold and b) with tile samples along the trajectory using IR-tiles
        (note manifold is different to \autoref{fig:embedding-manifold} which
        is created from model trained on RGB-tiles).
    }
    \label{fig:isomap-2nd-feb-trajectory}
\end{figure}

In the final use of the embedding manifold we demonstrate how it can be
used to study the temporal evolution of cloud organisation.
By applying the trained neural network on tiles sampled along a trajectory that
follow clouds as they evolve, we can map out transitions between organisational
states and thereby investigate the drivers behind these transitions.
We use a 4-day trajectory \citep[created with \textit{lagtraj},][]{lagtraj}, which
follows a cloud-layer airmass across the Atlantic as organisation develops from
initial isolated scattered cumuli (\textit{Sugar})
into larger isolated cumuli (\textit{Flowers}) \citep[August 2nd 2020, studied
by][]{Saffin_2023,Narenpitak_2021}. Tiles along the trajectory
(examples in \autoref{fig:isomap-2nd-feb-trajectory-samples}) are then mapped
onto the embedding manifold (\autoref{fig:isomap-2nd-feb-trajectory-manifold}).
The evolution on the manifold clearly shows, as is also indicated by the tile
samples shown, that over the first three days organisation exhibits diurnal
cycling that is eventually broken overnight ending in a drastically
different regime on February 2\textsuperscript{nd}.
In future work this form of manifold trajectory analysis will be used to unpick
the mechanisms behind this bifurcation behaviour.

\section{Conclusions}

In this work we have demonstrated that a map of all possible states of mesoscale
organisation exists as an emergent property of the internal embedding space
used by an unsupervised neural network, trained to group tiles of similar cloud
patterns together.

Examining this embedding manifold map we find that across the manifold the
visual variation in cloud patterns matches values of traditional metrics used
to measure clouds.
Although we find that traditional metrics used for measuring organisation are
able to capture the continuum of organisation variation, their varying
co-linearity may make them challenging to use in isolation to understand processes
of mesoscale organisation.
We find that the unsupervised
neural network "rediscovers" three (\textit{Sugar}, \textit{Flowers} and
\textit{Gravel}) of the four classes of organisation defined by
\cite{Stevens2019}, demonstrated by well-separated peaks in their manifold
distribution.
However, rather than appearing as isolated regions, these classes appear as
useful waymarkers in the full map of cloud organisation produced in this work.
By compositing ERA5 reanalysis onto the embedding manifold we find broad
agreement with prior work in stronger winds and lower sea-surface temperature
being associated with more cellular organisation. However, we also find that
the larger isolated clouds are principally found in conditions with high
cloud-level and low sub-cloud moisture (rather than lower tropospheric
stability being key).
Using the continuum representation of organisation we are able to show that
cloud organisation \textit{does} affect shortwave albedo beyond simply controlling
cloud-fraction. We find that, as expected, this is primarily due to changes
in the optical depth, the drivers of which can now be examined using the
manifold of organisation.
And finally, we have demonstrated how the ability to represent and measure the
continuum of organisation allows for the study of how convective organisation
develops; examining the evolution of \textit{Sugar} (scattered small cumuli)
to \textit{Flowers} (larger isolated cumuli) and capturing the breakaway from
diurnal cycling in organisation.

With the urgency of increased capacity to model Earth's future climate, this
novel technique to produce a \textit{map} of all states of convective
organisationa provides a new avenue for how to understand the processes of
cloud organisation, both in models and observations, and paves the way for
better representation in simulations of Earth's climate.

\acknowledgments
 The author acknowledges funding from the Paracon GENESIS (NERC NE/N013840/1) and EUREC\textsuperscript{4}A-UK (NERC NE/S015868/1) projects.
 GOES-16 data is available on the Amazon Open Data Registry at https://registry.opendata.aws/noaa-goes/.
 The Synoptic Radiative Fluxes and Clouds data sets
 (SYN1deg-1Hour\_Terra-Aqua,
 Edition 4A, https://doi.org/10.5067/TERRA+AQUA/CERES/SYN1DEG-1HOUR\_L3.004A and
 CER\_GEO\_Ed4\_GOE16\_NH\_V01.2 https://doi.org/10.5067/GOES16/CERES/GEO\_ED4\_NH\_V01.2) are made available by the NASA CERES group.

\bibliography{library.bib}






\end{document}